\documentclass[aps,prl,10pt,twocolumn,superscriptaddress,floatfix]{revtex4-2}
\usepackage{amsmath,amssymb,graphicx}
\usepackage[hidelinks]{hyperref}

\graphicspath{{./}}
\begin{document}
\title{Sandpile Topplings Generate a Monge--Ampère Field with Emergent Affine Geometry}
\author{Higinio Serrano}
\affiliation{Institute of Mathematics and Informatics, Bulgarian Academy of Sciences (IMI-BAS), Sofia, Bulgaria}
\author{Mikhail Shkolnikov}
\affiliation{Institute of Mathematics and Informatics, Bulgarian Academy of Sciences (IMI-BAS), Sofia, Bulgaria}
\author{Alexander Varypaev}
\affiliation{Institute of Mathematics and Informatics, Bulgarian Academy of Sciences (IMI-BAS), Sofia, Bulgaria}
\author{Nikita Kalinin}
\affiliation{Guangdong Technion--Israel Institute of Technology (GTIIT), Shantou, China}
\author{Ernesto Lupercio}
\affiliation{Institute of Mathematics and Informatics, Bulgarian Academy of Sciences (IMI-BAS), Sofia, Bulgaria}
\affiliation{Departamento de Matem\'aticas, CINVESTAV, Ciudad de M\'exico, Mexico}

\begin{abstract}
One grain can trigger an avalanche in an Abelian sandpile prepared at three grains per site. After many dilute additions, the rescaled toppling counts approach a scalar field. The nonlinear Monge--Ampère equation fixes its shape and amplitude. Grains move in four directions, yet the field follows area-preserving stretches and shears of the pile and its drive. Doubling a uniformly driven disk’s width and halving its height should preserve the peak of the mean toppling profile. A Poisson law calibrated on the disk predicts a drop to $47\%$. In simulations the ellipse retains $98.6\%$ of the disk peak.
\end{abstract}
\maketitle

The Abelian sandpile is a paradigmatic model of self-organized criticality in which the slow addition of grains and their redistribution by local topplings generate avalanches over a broad range of scales~\cite{BTW1987,Dhar1990}. We study the nonstationary response of a maximally stable pile, with three grains per site. The first avalanche on a square reaches every site. The odometer counts how often each topples. On a square with lattice spacing $h=1/1600$, $10\,192$ additions produced $3.25\times10^{11}$ topplings per pile on average. A few additions leave lines of depleted sites that trace a tropical curve~\cite{KalininShkolnikov2016,Kalinin2018PNAS}. Lupercio and Shkolnikov~\cite{LS2026} conjectured that after many dilute additions, the rescaled toppling counts approach a single scalar field governed by the nonlinear Monge--Ampère equation, with a continuous affine symmetry absent from the lattice. We test the law on a square and its symmetry by stretching a disk and its drive into an ellipse.

Double the width of a uniformly driven disk and halve its height, carrying the sources along. The nearest escape boundary is now half as far from the center. Each toppling still sends two grains vertically and two horizontally. Should the center still topple as often? Both piles have the same uniform density $\rho$, so any common law $-\Delta F=g(\rho)$ fixes the same sum of curvatures. The peak falls to $8/17$, or $47\%$, of its disk value, even for $g(\rho)=2\sqrt{\rho}$, which matches the entire disk profile. The Monge--Ampère equation for the limiting tropical field, $\det D^2F=\rho$~\cite{KLSS2026}, fixes the product of the curvatures. Steepening one direction can then compensate exactly for flattening the other. The continuum law predicts that every contour of a given activity stretches with the disk. The peak therefore stays fixed.

We simulate the Abelian sandpile on a bounded convex region $\Omega$ cut from a square lattice of spacing $h$. Each site starts with three grains. Growth from a single source on an empty lattice has a different continuum limit governed by an obstacle problem~\cite{PegdenSmart2013}. We add $N$ grains at positions sampled independently from a probability density $\rho$ of unit integral. Whenever a site holds at least four grains, it topples, sending one grain to each neighbor. Grains that cross the boundary leave the pile. We may let each avalanche finish before adding the next grain or add all the grains at once. For the same landing sites the Abelian property gives the same toppling count $H_{h,N}(x)$ at every site $x$. To compare piles at different lattice spacings and drive strengths we use the rescaled odometer
\begin{equation}
 U_{h,N}(x)=\frac{h}{\sqrt N}\,H_{h,N}(x).
 \label{eq:odo}
\end{equation}
We take $h\to0$ and $N\to\infty$ with $Nh^2\to0$. A patch of fixed area within the drive receives many grains. A single site near $x$ receives only $Nh^2\rho(x)$ additions on average, to leading order. Its activity comes chiefly from grains that landed elsewhere.

\begin{figure*}[t]
\centering\includegraphics[width=\textwidth]{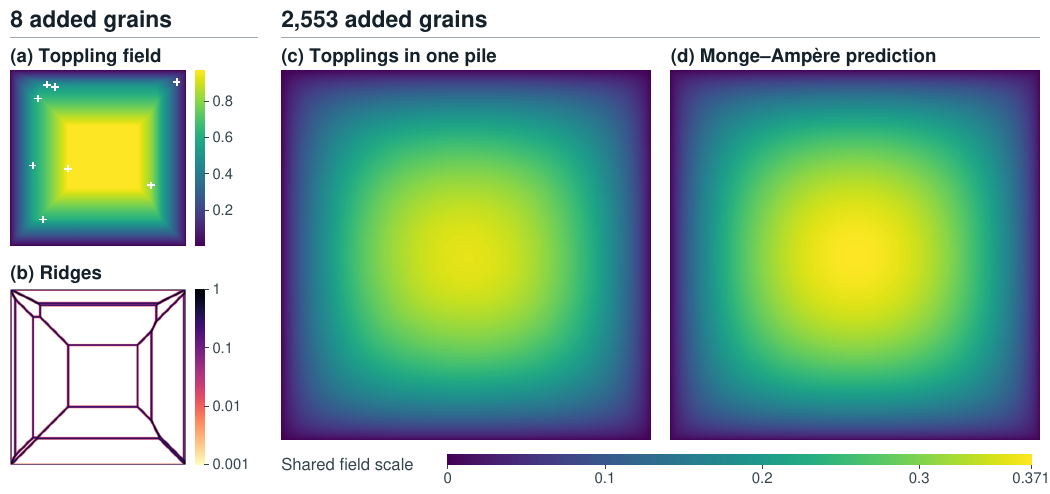}
\caption{Topplings on two uniformly driven squares at $h=1/400$. (a) The profile $hH$ after eight additions, where $H$ counts topplings at a site. Crosses mark the landing sites. (b) Magnitude of the lattice Laplacian of $hH$ after Gaussian smoothing over two lattice spacings, relative to its maximum on a logarithmic scale. The ridges separate adjacent facets. (c) The rescaled count $U=hH/\sqrt N$ for a single pile with 2553 additions. (d) The independently calculated Monge--Ampère profile for the same square and uniform drive, plotted on the scale of (c). Plotting code was AI-assisted.}
\label{fig:fields}
\end{figure*}

In the simulated square of Fig.~\ref{fig:fields}(c), fewer than one site in 250 receives a grain directly, yet the average site topples nearly four thousand times. Beside it is a field we can calculate before dropping the first grain [Fig.~\ref{fig:fields}(d)]. We know the shape of the square, that activity vanishes at its boundary, and that equal areas are equally likely to receive a grain. The actual landing sites enter only the simulation. Plotted on the same color scale, both fields rise to a broad maximum near the center. Farther out, their contours flatten along the four sides (Supplemental Material~\cite{SM}, Sec.~S4).

The random landing sites matter even when we refine the lattice. On a uniformly driven disk, halving the lattice spacing at fixed $N$ scarcely changes the typical discrepancy between a single pile and the continuum prediction. The finer grid resolves each avalanche in greater detail, but the number of grains in a given patch still fluctuates from pile to pile. More additions sample the drive more evenly. On a fixed lattice, however, a sufficiently heavy drive moves even the mean profile farther from the prediction (Supplemental Material~\cite{SM}, Sec.~S4). A dilute drive leaves room between neighboring sources. The eight-grain pile in Fig.~\ref{fig:fields}(a,b) shows what forms in that space.

Between the landing sites the toppling profile $hH$ rises across flat facets. In the tropical description each facet has two integer slope components, and the landing sites lie on ridges where the slope changes~\cite{KalininShkolnikov2016,Kalinin2018PNAS}. Mark each slope as a point in a plane. Across a ridge two such points span a segment. Around an elementary junction three slopes enclose a triangle of area $1/2$. A generic ridge network has one bounded cell per added grain and, up to a boundary correction, two three-way junctions per cell. Its total slope area therefore grows as $N$~\cite{LS2026,KLSS2026}. Doubling the height of a profile doubles both components of every slope and quadruples that area. This fixes the $\sqrt N$ height scale in the continuum description proposed for the sandpile~\cite{LS2026}. After rescaling, the slope area associated with a patch approaches the fraction of grains that landed there~\cite{KLSS2026}. Where the limiting field $F$ is smooth, $D^2F$ maps a displacement on the pile to a change of slope. Its determinant measures how much slope area a unit area of the pile supplies. Thus the density of additions selects the concave profile through
\begin{equation}
 \det D^2F=\rho\quad\text{in }\Omega,\qquad F=0\quad\text{on }\partial\Omega.
 \label{eq:law}
\end{equation}
For generic sources bounded away from a rational-slope polygon’s boundary, the lattice limit follows on meshes and sites chosen to resolve each cloud’s tropical profile~\cite{KLSS2026}. In the simulations we refine the lattice as we increase the number of sources, and the random additions can reach straight or curved boundaries.

A grain crossing a bond suggests a different continuum law. Each toppling at $x$ sends one grain to its neighbor $y$, and each toppling at $y$ sends one back. The net transfer from $x$ to $y$ is exactly $H(x)-H(y)$. Summing over the four bonds gives the net export from a site, the negative lattice Laplacian of $H$. In a stationary pile the mean height at each site is unchanged by a grain addition. Net export then balances deposition on average, and the mean response to a grain added at a given site is the lattice Green function~\cite{Dhar1990}. For the uniformly driven unit disk, the stationary response predicts $N/(4\pi)$ topplings at the center. The prepared-pile field predicts a larger count by a factor $2\sqrt{\pi}/\sqrt{Nh^2}$. The measured factor rises from $44.5$ at $h=1/800$ to $89.3$ at $1/3200$ along the paired-disk path (Supplemental Material~\cite{SM}, Sec.~S5). A site can finish with two grains, one fewer than it started with. It has then exported one grain more than it received in direct additions. Which sites supply these extra grains is itself an outcome of the relaxation. To see how much of the activity a linear law can nevertheless explain, we solve $-\Delta P=\kappa\rho$ with the same zero boundary value and allow the Poisson amplitude $\kappa$ to fit the pile.

The uniform disk seems an ideal place to decide between the two laws. Equation~(\ref{eq:law}) predicts the whole profile in absolute units, $F(r)=(1-r^2)/(2\sqrt{\pi})$ on the unit disk, and the simulated mean field closely follows this parabola. The relative interior discrepancy is $0.38\%$ at $h=1/3200$ on a dilution path selected in a coarser scan (Supplemental Material~\cite{SM}, Sec.~S4). Yet matching the peak makes the Poisson field agree with the entire parabola as well. The two principal curvatures are equal and constant. Monge--Ampère fixes their product. A single choice of Poisson amplitude reproduces their sum everywhere on the disk. Concentrating the drive near the center breaks this coincidence. Inside radius $r$, let $M(r)$ be the fraction of all additions. The slopes fill a disk of area $\pi |F'(r)|^2=M(r)$, whereas the Poisson flux through the circle gives $2\pi r|P'(r)|=\kappa M(r)$. Once most of the drive is enclosed, the sandpile law predicts a nearly constant radial slope and the Poisson slope falls approximately as $1/r$. With a centered Gaussian drive the simulated profile follows the Monge--Ampère prediction more closely, even after the Poisson amplitude is fitted (Supplemental Material~\cite{SM}, Sec.~S4).

\begin{figure}[ht]
\centering\includegraphics[width=\columnwidth]{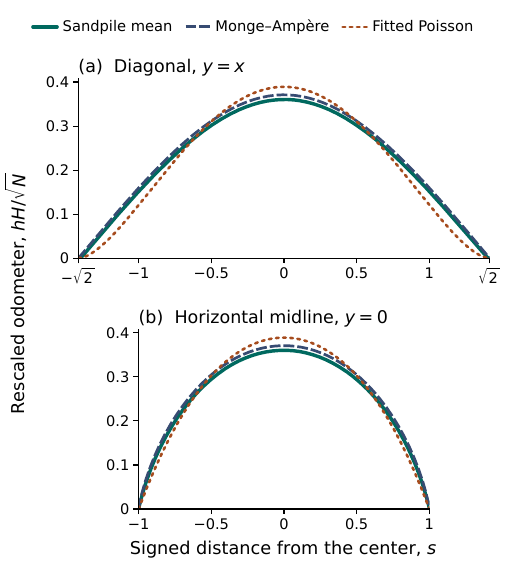}
\caption{Activity on the uniformly driven square along (a) a diagonal and (b) the horizontal midline. Both show signed distance from the center. Curves average 20 piles at $h=1/800$ with 5099 additions each. A single Poisson amplitude fits the interior mean by least squares. The drive fixes the Monge--Ampère amplitude. The quoted interior errors use 20 separate piles at $h=1/3200$ with 6113 additions each. Plotting code was AI-assisted.}
\label{fig:square}
\end{figure}

On the square the boundary alone separates the two laws. Restore uniform drive and follow the activity from the center to a side or toward a corner [Fig.~\ref{fig:square}]. Give Poisson the amplitude that best fits the mean field over the entire interior. Near the center its curve lies above the data. Along the diagonal it crosses the measured profile and falls below it. Lowering the amplitude to bring the peak into agreement would leave even fewer topplings on the approach to the corner. The Monge--Ampère field follows both cuts on the absolute scale fixed by the drive. At $h=1/3200$ the relative interior $L^2$ errors are $1.12\%$ for Monge--Ampère and $9.58\%$ for fitted Poisson. This square loading was selected from a scan at $h=1/800$. In a separate test we combined two loadings to estimate the dilute profile, with extrapolation bounds set before the finer runs. Its discrepancy from Monge--Ampère falls below the estimated sampling noise. The Poisson discrepancy persists even after refitting (Supplemental Material~\cite{SM}, Sec.~S4).

\begin{figure*}[t]
\centering\includegraphics[width=0.96\textwidth]{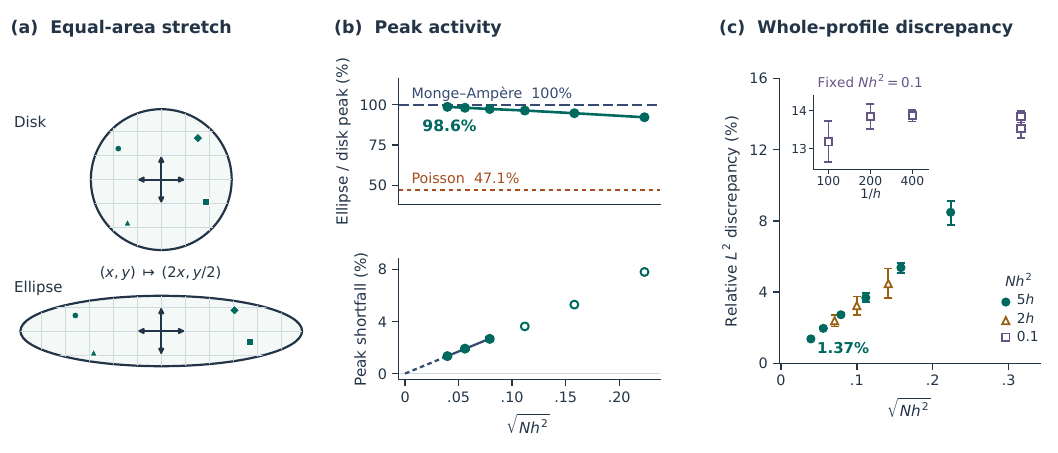}
\caption{Symmetry test on the uniformly driven disk and ellipse. (a) The stretch carries the boundary and sources together. Toppling directions stay fixed. (b) Peak ratio of the mean ellipse and disk fields (top) and its shortfall from unity (bottom) versus $\sqrt{Nh^2}$. Below, the solid line fits the three finest meshes (filled points) with a free intercept. Dashes extrapolate to zero loading. (c) Whole-profile discrepancy, normalized by the ellipse interior $L^2$ norm and corrected for sampling variance. Circles and triangles follow $Nh^2=5h$ and $2h$. The inset resolves the fixed-loading squares by mesh. Each point uses 20 pairs. Bars give paired 95\% bootstrap intervals in (c) and at the three finest meshes in (b). Plotting code was AI-assisted.}
\label{fig:ellipse}
\end{figure*}

Once the square profile is known, the nonlinear law makes a prediction for a different pile. Shear the square into a parallelogram by sending $(x,y)$ to $(x+y,y)$ and carry every source along. A vertical bond would shear into a diagonal, yet the topplings in the new pile still send grains along the original axes. The measured field follows the shear increasingly closely as we refine the lattice (Supplemental Material~\cite{SM}, Sec.~S5). At the scale of the facets the shear is already a symmetry. It takes a slope $(p,q)$ to $(p,q-p)$, so all integer slopes remain integer~\cite{LS2026,KLSS2026}. The same group $SL(2,\mathbb Z)$ enters the conservation laws for sandpile strings~\cite{Caracciolo2010}. The stretch from the disk to the ellipse takes the unit slope $(1,0)$ to $(1/2,0)$ and lies outside this group. What allows the field to follow it? The rescaled tropical slopes are spaced only $1/\sqrt N$ apart. As they fill the slope plane more densely, what the drive fixes in each patch is an area. An area-preserving map $A$ carries the patch and its sources together, while the slopes transform by $A^{-T}$, which preserves their area as well. Thus the transported field has the correct slope area everywhere and vanishes on the new boundary. This affine covariance preserves the field height at corresponding points for every $A\in SL(2,\mathbb R)$.

At $h=1/3200$, with $16\,000$ additions per pile, the peak of the mean ellipse field retains $98.6\%$ of the disk value [Fig.~\ref{fig:ellipse}(b)]. Poisson predicts $47\%$. Across six meshes on the path $Nh^2=5h$, the peak shortfall is approximately $0.34\sqrt{Nh^2}$. Extrapolating the three finest ratios linearly in $\sqrt{Nh^2}$ gives $(99.98\pm0.07)\%$ at zero loading (one sampling standard error). Affine covariance asks more of the pile than agreement at the peak. Take two points at equal radius on the two axes of the disk. After the stretch one lies four times as far from the center as the other, yet the field assigns them the same activity. We carry each grain addition from the disk to the ellipse and let both piles relax along the original lattice axes. Across the interior, the discrepancy between the simulated ellipse field and the stretched disk field falls along both dilute paths [Fig.~\ref{fig:ellipse}(c)]. Refining at fixed loading leaves a discrepancy near $14\%$. In the two finest ensembles, the peak ratios and whole-profile discrepancies agree with extrapolations from the coarser meshes (Supplemental Material~\cite{SM}, Sec.~S5).

If the activity follows the stretch, what happens to the grains left behind? Draw a window $W$ in the pile and add up the bond currents around it. Every transfer internal to the window cancels. Across each boundary bond the difference in toppling counts, inside minus outside, gives the net number of grains that crossed outward. Subtract the grains deposited in the window and what remains is exactly the number missing from its initial stock. This balance uses only the accumulated counts (Supplemental Material~\cite{SM}, Sec.~S6). Substituting the continuum field for the odometer gives a prediction for the mean deficit per site,
\begin{equation}
 \bar\delta(W)\simeq h\sqrt{N}\,\langle-\Delta F\rangle_W
                  -Nh^2\langle\rho\rangle_W ,
 \label{eq:deficit}
\end{equation}
where brackets denote window averages. In the interior of a uniformly driven square the field accounts closely for the missing grains. Near a straight side the continuum profile has an infinite inward slope at the boundary. For windows near straight sides, a fit of measured to predicted losses gives a slope of about one half (Supplemental Material~\cite{SM}, Sec.~S6). Grain balance remains exact. The difficulty is estimating the boundary current from the continuum field on a finite lattice.

We can count the missing grains in the same disk and ellipse that gave the peak measurement. Take the central quarter of the disk by area and stretch that window with the pile. On average the two windows receive nearly the same share of the drive. Both export many more grains than land in them, drawing on grains already in the pile before the first avalanche. The stretch that barely changes the peak activity more than doubles the bulk deficit. On the finest paired lattice, the ratio of deficits is $2.107$ and Eq.~(\ref{eq:deficit}) predicts $2.138$. The prediction lies above the paired sampling interval on all three fine meshes, but the gap narrows as the loading falls (Supplemental Material~\cite{SM}, Sec.~S6).

On the disk the toppling profile bends equally along both axes. Carry it to the ellipse at fixed peak height. The short-axis curvature grows fourfold in magnitude and the long-axis curvature falls to one quarter of its disk value. Their product stays fixed and the profile vanishes on the new boundary, so it solves Eq.~(\ref{eq:law}) for the same uniform drive. The sum of their magnitudes, $-\Delta F$, sets the net outward transport and rises to $17/8$ of its disk value. Direct deposition in Eq.~(\ref{eq:deficit}) falls as $\sqrt{Nh^2}$ relative to outward transport, so the dilute deficit ratio should also tend to $17/8$. Assuming a leading correction proportional to $\sqrt{Nh^2}$, the three finest meshes give $2.125\pm0.006$ at zero loading (one sampling standard error, Supplemental Material~\cite{SM}, Sec.~S6). The center can topple just as often and still receive fewer grains back from its neighbors.

Every grain in these simulations moves along the same four lattice directions, even when the boundary and the sources are stretched. What acquires the continuous symmetry is the count of topplings accumulated over many avalanches. In a pile prepared at three grains per site, one scalar field governed by the Monge--Ampère equation describes that activity and, through grain conservation, the depletion it leaves behind. To test how much the symmetry depends on this preparation, we could remove one grain from each of a small fraction of randomly chosen sites before adding the first grain. Does the emergent affine symmetry survive these dilute defects? Would an equal-area stretch of the pile and its drive still stretch the activity contours while leaving the peak unchanged?

\begin{acknowledgments}
\textit{Acknowledgments.}
The work of E.L., H.S., and M.S. was supported by the Simons Foundation under grant SFI-MPS-T-Institutes-00007697 and by the Ministry of Education and Science of the Republic of Bulgaria under grant DO1-239/10.12.2024. E.L. acknowledges Cinvestav for sabbatical leave during which most of this work was performed, and the Institute for the Mathematical Sciences of the Americas (IMSA) at the University of Miami for hospitality and support during the preparation of this work.
\end{acknowledgments}

\section*{Data Availability}
The numerical tables, random seeds, simulation code, figure inputs, saved fields and numerical reference solutions used for the reported comparisons are available from the corresponding author, Ernesto Lupercio, on request. The Supplemental Material~\cite{SM} describes the calculations and validation.

\section*{Use of generative AI tools}
The physical ideas, the choice of observables and the conclusions are the authors' own. Claude Opus 5.5 (Anthropic) agents assisted with simulation and plotting code, production runs, analysis of results and drafting. OpenAI Codex (GPT-6.2) assisted with the paired disk--ellipse analysis, figure preparation, consistency checks and revision. ChatGPT with GPT-6 Astra (OpenAI) assisted with concept and reference queries. Gemini 4 (Google) assisted with concept queries. Claude Fable 5.2 (Anthropic) was used for pre-submission review; some analyses it suggested were adopted after the authors checked them. The authors verified the results and take responsibility for the manuscript (Supplemental Material~\cite{SM}, Sec.~S3).

\end{document}